# A pseudobinary approach in multicomponent interdiffusion

Aloke Paul,
Dept. of Materials Engineering, Indian Institute of Science, Bangalore, India
E-mail: aloke@materials.iisc.ernet.in, Tel.: 918022933242, Fax: 91802360 0472

**Abstract**

Interdiffusion studies become increasingly difficult to perform with the increasing number of elements in a system. It is rather easy to calculate the interdiffusion coefficients for all the compositions in the interdiffusion zone in a binary system. The intrinsic diffusion coefficients can be calculated for the composition of Kirkendall marker plane in a binary system. In a ternary system, however, the interdiffusion coefficients can only be calculated for the composition where composition profiles from two different diffusion couples intersect. Intrinsic diffusion coefficients are possible to calculate when the Kirkendall markers are also present at that composition, which is a condition that is generally difficult to satisfy. In a quaternary system, the composition profiles for three different diffusion couples must intersect at one particular composition to calculate the diffusion parameters, which is a condition that is almost impossible to satisfy. To avoid these complications in a multicomponent system, the average interdiffusion coefficients are calculated. I propose a method of calculating the intrinsic diffusion coefficients and the variation in the interdiffusion coefficients for multicomponent systems. This method can be used for a single diffusion couple in a multicomponent pseudobinary system. The compositions of the end members of a diffusion couple should be selected such that only two elements diffuse into the interdiffusion zone. A few hypothetical diffusion couples are considered in order to validate and explain our method. Various sources of error in the calculations are also discussed.

## 1. Introduction

Diffusion is one of the most important phenomena to study because it controls many physical and mechanical properties of materials, such as microstructural stability, creep, growth of precipitates, and the kinds of product phases that form between dissimilar materials. In many products such as coatings, flip chips, and wire bonding in electronic packaging, the performance of a structure depends on the diffusion-controlled growth of brittle intermetallic compounds. Some products such as $Nb_3Sn$, $V_3Ga$, $V_3Si$ intermetallic superconductors, bond coats in turbine blades, diffusion bonds, and laminate structures are produced during diffusion. The tracer method and diffusion coupling are the two main methods of studying solid-state diffusion. The former is a more fundamental method and is very useful for understanding the physics; that is, the atomic mechanism, of diffusion. The latter is similar to practical applications of diffusion in which the diffusion is controlled by the composition or the activity gradient of the system. Using this method, we study interdiffusion of elements and calculate many important diffusion parameters such as interdiffusion and intrinsic diffusion coefficients, and one can even indirectly calculate the tracer diffusion coefficients [**1-7**] or the ratio of the tracer diffusion coefficients [**8-17**].

However, the limitation of using this method to study interdiffusion is that we cannot study multicomponent systems to calculate the diffusion parameters [**18**], although many elements are used in most of the systems to balance the properties. It is rather straightforward to study binary systems, and only a single experiment is required in order to calculate the interdiffusion coefficients for all compositions in the interdiffusion zone. This is why most studies are performed on binary systems. One can calculate the intrinsic diffusion coefficients for the Kirkendall marker plane if inert particles are used to locate it [**19-26**]. Ternary systems, on the other hand, are much more complicated to study than binary ones. One can only calculate the main and cross interdiffusion coefficients for the point of intersection of the composition profiles for two different diffusion couples [27]; therefore, many experiments are required in order to calculate the diffusion parameters for only a few compositions of elements in a system. Calculating the intrinsic diffusion coefficients is even more difficult since not only two composition profiles should intersect at one particular composition but also the Kirkendall marker plane must be present for that composition. Finding the inert marker plane at the point of intersection is mostly a matter of luck. This is why there are very few studies on ternary systems and why the intrinsic diffusion coefficients are rarely calculated in those studies [27]. Interdiffusion studies are also available on quaternary systems [28, 29], and this will be explained in more detail later. Not all the main and cross interdiffusion coefficients were calculated for the quaternary systems because calculating them would require three different diffusion couples to intersect at one common composition, and it is very difficult to design experiments that fulfil this condition. Therefore, only the main interdiffusion coefficients were calculated for certain diffusion couples in the Ni-Cr-Co-Mo system when two elements did not diffuse (i.e. did not develop a concentration profile). In many cases, the average effective interdiffusion coefficients are calculated in multicomponent systems [30]. Krishtal *et al*. [31] considered constant diffusion coefficients to develop a method of calculating the diffusion parameters for multicomponent systems. Thompson and Morral [32] subsequently considered constant diffusion coefficients to develop a square root of diffusivity approach for multicomponent systems. These methods can be used when the difference in the compositions of the diffusion couples is sufficiently small such that the diffusion coefficients do not significantly vary over the range of compositions. The method developed by Thompson and Morral was used to calculate the diffusion parameters for a few quaternary systems by maintaining a very small difference in the compositions of the end members of a diffusion couple, i.e. at 5 at% [33-35]. Stalker *et al*. [36] showed that the interdiffusion flux could be back-calculated with very little error when the compositions of the end members of the diffusion couple in a quaternary system varied over such a small range. However, it is not possible to obtain much relevant information from these studies since calculating interdiffusion coefficients can be sensitive to the compositions of the end components of diffusion couples. Therefore, the aim of this study is to develop a method of calculating the variation in diffusion parameters and the Kirkendall effect for multicomponent systems.

## 2. Background of the problem

The interdiffusion flux of an *i*th element in a multicomponent system can be written as [30]

$$\tilde{J}_i = -\sum_{j=1}^{n-1} \tilde{D}_{ij}^n \frac{\partial C_j}{\partial x}, \tag{1}$$

where $\tilde{J}_i$ is the interdiffusion flux of element *i*, $\tilde{D}$ is the interdiffusion coefficient, $C$ is the concentration, $x$ is the position parameter, $n$ is the dependent variable, such that ($n$-1) interdiffusion coefficients are required to determine the interdiffusion fluxes in a multicomponent system.

The interdiffusion flux of an element is related to the intrinsic flux of elements by [30]

$$\tilde{J}_i = J_i - N_i \sum_{k=1}^{n} J_k, \tag{2}$$

where $J_i$ is the intrinsic flux of element *i*.

In a binary system, this leads to

$$\tilde{J}_1 = -\tilde{D}\frac{dC_1}{dx} \tag{3a}$$

$$\tilde{J}_2 = -\tilde{D}\frac{dC_2}{dx} \tag{3b}$$

The relations above indicate that there is only one interdiffusion coefficient required to explain the interdiffusion flux calculated based on any of the elements. Using the standard thermodynamic relation $V_1 dC_1 + V_2 dC_2 = 0$, it can be shown that [19]

$$V_1 \tilde{J}_1 + V_2 \tilde{J}_2 = 0, \tag{4a}$$

where $V_i$ is the partial molar volume of element $i$.

In many systems, the variation of molar volume with composition is not known and considering constant molar volume, **Eq. 4a** can be written as

$$\tilde{J}_1 + \tilde{J}_2 = 0, \tag{4b}$$

Intrinsic and interdiffusion fluxes in a binary system from Eq. 2 can be written as

$$\tilde{J}_1 = J_1 - N_1(J_1 + J_2) \Rightarrow \tilde{J}_1 = (1 - N_1)J_1 - N_1 J_2 \Rightarrow N_2 J_1 - N_1 J_2 \tag{5a}$$

$$\tilde{J}_2 = J_2 - N_2(J_1 + J_2) \Rightarrow \tilde{J}_1 = (1 - N_2)J_2 - N_2 J_1 \Rightarrow N_1 J_2 - N_2 J_1 \tag{5b}$$

since $N_1 + N_2 = 1$.

In multicomponent systems, the intrinsic flux is related to the intrinsic diffusion coefficients by [30]

$$J_i = -\sum_j^{n-1} D_{ij}^n \frac{\partial C_j}{\partial x}. \tag{6a}$$

In a binary system, intrinsic fluxes are related to the intrinsic diffusion coefficient, $D_i$ by

$$J_1 = -D_1 \frac{dC_1}{dx} \tag{6b}$$

$$J_2 = -D_2 \frac{dC_2}{dx} \tag{6c}$$

Both the intrinsic fluxes are directly related to the interdiffusion flux and since the interdiffusion fluxes are the same for both the elements with opposite sign, one can determine the intrinsic fluxes or intrinsic diffusion coefficients from any of the composition profiles at the Kirkendall marker plane [30].

In a ternary system ($n = 3$), $\tilde{J}_1$ and $\tilde{J}_2$ can be expressed as

$$\tilde{J}_1 = -\tilde{D}_{11}^3 \frac{\partial C_1}{\partial x} - \tilde{D}_{12}^3 \frac{\partial C_2}{\partial x} = -\tilde{D}_{11}^3 \frac{1}{V_m} \frac{\partial N_1}{\partial x} - \tilde{D}_{12}^3 \frac{1}{V_m} \frac{\partial N_2}{\partial x} \tag{7a}$$

$$\tilde{J}_2 = -\tilde{D}_{21}^3 \frac{\partial C_1}{\partial x} - \tilde{D}_{22}^3 \frac{\partial C_2}{\partial x} = -\tilde{D}_{21}^3 \frac{1}{V_m} \frac{\partial N_1}{\partial x} - \tilde{D}_{22}^3 \frac{1}{V_m} \frac{\partial N_2}{\partial x} \tag{7b}$$

$\tilde{J}_3$ can be related by

$$\tilde{J}_1 + \tilde{J}_2 + \tilde{J}_3 = 0. \tag{7c}$$

The relations above are true for a constant molar volume, $V_m$. In most of the ternary systems, the change in lattice parameter with composition is not known and an average molar volume is considered for calculations. In Eqns. **7a** and **7b**, $\tilde{D}_{11}^3$ and $\tilde{D}_{22}^3$ are the main or direct interdiffusion coefficients, which represent the influence of concentration gradient of one element on the diffusion rate of the same element. $\tilde{D}_{12}^3$ and $\tilde{D}_{21}^3$ are the cross or indirect interdiffusion coefficients, which represent the influence of concentration gradient of one element on the diffusion rate of other element. Element 3 is the dependent variable. $C_i = \frac{N_i}{V_m}$ is the concentration and $N_i$ is composition in mol or atomic fraction.

Therefore, the values of $\tilde{D}_{11}^3, \tilde{D}_{12}^3, \tilde{D}_{21}^3$ and $\tilde{D}_{22}^3$ are required to explain the interdiffusion flux in a ternary system. However, the calculation of these parameters is not as straightforward as in a binary system. From **Eq. 7**, it must be clear that these diffusion parameters can be calculated at the intersection of two diffusion profiles from two different diffusion couples with different end members. Interdiffusion fluxes are related to the intrinsic fluxes by

$$\tilde{J}_1 = (1 - N_1)J_1 - N_1(J_2 + J_3) \tag{8a}$$

$$\tilde{J}_2 = (1-N_2)J_2 - N_2(J_1 + J_3) \tag{8b}$$

$$\tilde{J}_3 = (1-N_3)J_3 - N_3(J_1 + J_2) \tag{8c}$$

In a ternary system these fluxes in terms of intrinsic diffusion coefficients can be written as [**30**]

$$\begin{aligned} J_1 &= -D_{11}^3 \frac{\partial C_1}{\partial x} - D_{12}^3 \frac{\partial C_2}{\partial x} \\ J_2 &= -D_{21}^3 \frac{\partial C_1}{\partial x} - D_{22}^3 \frac{\partial C_2}{\partial x} \\ J_3 &= -D_{31}^3 \frac{\partial C_1}{\partial x} - D_{32}^3 \frac{\partial C_2}{\partial x} \end{aligned} \tag{9}$$

That means the calculation of the intrinsic diffusion coefficients is very difficult since not only two composition profiles should intersect at one particular composition of interest, even the markers also should be present at that composition. Ideally, there is a little chance.

In a quaternary system, the relations become even more complicated. Minimum three interdiffusion fluxes are required and each of the interdiffusion flux is related to three interdiffusion coefficients. For example, interdiffusion flux of element 1 can be expressed as

$$\tilde{J}_1 = -\tilde{D}_{11}^4 \frac{\partial C_1}{\partial x} - \tilde{D}_{21}^4 \frac{\partial C_2}{\partial x} - \tilde{D}_{31}^4 \frac{\partial C_3}{\partial x} \tag{10}$$

So there are nine interdiffusion coefficients to determine and hence we need three composition profiles to intersect at one particular composition, which is almost impossible to get. To circumvent this problem, in general, average effective diffusion coefficients for each of the elements are determined. If there is significant concentration change in the interdiffusion zone, it is not possible to determine the concentration dependence of the interdiffusion coefficients. As explained already, it is also extremely difficult to calculate the intrinsic diffusion coefficients when more than two species are present in the system. So the aim of the manuscript is to develop a relatively simple approach to study interdiffusion in multicomponent systems. Following, the restrictions of this approach are discussed.

## 3. Pseudobinary approach to study interdiffusion in multicomponent systems

### 3.1 Calculation of diffusion parameters in a binary system

As discussed in the previous section, it is rather simple to study interdiffusion and calculate all the important diffusion parameters in a binary system. But as we increase the number of elements, it becomes very difficult to impossible. Before explaining the new approach, let me review the experimental procedure and calculation of diffusion parameters in a binary system so that the approach used in multicomponent systems can be understood. A diffusion couple of two blocks with compositions, let say alloy 1: $A_{0.15}B_{0.85}$ and alloy 2: $A_{0.35}B_{0.65}$, as shown in **Fig. 1a**. Suppose in the middle, one phase, γ grows with a composition range of $\Delta N_A^\gamma$ = 0.2-0.3. For the sake of simplicity, we consider a linear change in composition in the interdiffusion zone, as shown in composition profiles of A and B in **Figs. 1b and c,** respectively. Further, we consider that there is no change in molar volume with the change in composition. The assumptions are taken such that it will be easier for the readers to calculate the diffusion parameters without the help of any software. It should be noted here that this method is applicable in the real systems, where composition profile is not linear and the molar volume varies with composition. Further, we consider the presence of the marker plane to calculate the intrinsic diffusivities at that plane (or composition). Interdiffusion and intrinsic diffusion coefficients at that plane are only calculated so that we can validate our results. Interdiffusion flux of an element, *i* in binary or multicomponent systems can be calculated by [**30**]

$$\tilde{J}_i = -\frac{\Delta N_i}{2tV_m^\gamma}\left[(1-Y_i)\int_{x^{-\infty}}^{x^*} Y_i dx + Y_i \int_{x^*}^{x^{+\infty}}(1-Y_i)dx\right] \tag{11a}$$

We can calculate the interdiffusion flux of A as

$$\tilde{J}_A = -\frac{\Delta N_A}{2tV_m^\gamma}\left[(1-Y_A)\int_{x^{-\infty}}^{x^*} Y_A dx + Y_A \int_{x^*}^{x^{+\infty}}(1-Y_A)dx\right], \tag{11b}$$

where minus sign comes from the fact that element A diffuses from right to left. For the interdiffusion flux of B, minus sign will not be there. $\Delta N_A = N_A^+ - N_A^-$, $N_A^+$ and $N_A^-$ are the composition of the initial and unaffected parts of the left and right hand side of the end members. $x^*$ is the position of interest. $Y_A$ is the composition normalized variable and equal to $\frac{N_A - N_A^-}{N_A^+ - N_A^-}$. From **Eq. 3** for constant molar volume, we can write

$$\tilde{J}_A = -\tilde{D}\frac{dC_A}{dx} = -\tilde{D}\frac{1}{V_m^\gamma}\frac{dN_A}{dx} \tag{11c}$$

From **Eq. 11b** and **c**, we can write the interdiffusion coefficient at the marker plane (denoted by K) as

$$\begin{aligned}\tilde{D} &= \frac{1}{2t}\left(\frac{dx}{dN_A/\Delta N_A}\right)_K\left[(1-Y_A^K)\int_{x^{-\infty}}^{x_K} Y_A dx + Y_A^K \int_{x_K}^{x^{+\infty}}(1-Y_A)dx\right] \\ &= \frac{1}{2t}\left(\frac{dx}{dY_A}\right)_K\left[(1-Y_A^K)\int_{x^{-\infty}}^{x_K} Y_A dx + Y_A^K \int_{x_K}^{x^{+\infty}}(1-Y_A)dx\right]\end{aligned} \tag{11d}$$

It should be noted here that the data calculated will be the same if it is calculated with respect to composition profile of B instead of A. Let us consider that markers used at the interface before annealing are shifted to the location of 0.24 atomic fraction A (0.76 atomic fraction B) located at 40 μm from the alloy 1/γ phase interface. The total thickness of the product γ phase grown in the interdiffusion zone is 100 μm after 25 h of annealing at one particular temperature, T. Interdiffusion coefficients can be calculated for any of compositions; however, we calculate it at the marker plane to validate our results. With the help of **Eq. 11d** and from the $Y_A$ vs. $x$ plot, as shown in **Fig. 1d**, the interdiffusion coefficient is calculated as $2.05 \times 10^{-14}$ m$^2$/s.

Intrinsic diffusion coefficients of elements A and B can calculated from [19]

$$D_A = \frac{1}{2t}\left(\frac{dx}{dC_A}\right)_K \left[ N_A^+ \int_{x^{-\infty}}^{x_K} \frac{Y_A}{V_m} dx - N_A^- \int_{x_K}^{x^{+\infty}} \left(\frac{1-Y_A}{V_m}\right) dx \right] = \frac{1}{2t}\left(\frac{dx}{dN_A}\right)_K \left[ N_A^+ \int_{x^{-\infty}}^{x_K} Y_A dx - N_A^- \int_{x_K}^{x^{+\infty}} (1-Y_A) dx \right]$$

(12a)

$$D_B = \frac{1}{2t}\left(\frac{dx}{dC_B}\right)_K \left[ N_B^+ \int_{x^{-\infty}}^{x_K} \frac{Y_A}{V_m} dx - N_B^- \int_{x_K}^{x^{+\infty}} \left(\frac{1-Y_A}{V_m}\right) dx \right] = -\frac{1}{2t}\frac{dx}{dN_A}\left[ N_B^+ \int_{x^{-\infty}}^{x_K} Y_A dx - N_B^- \int_{x_K}^{x^{+\infty}} (1-Y_A) dx \right]$$

(12b)

since we are considering a constant molar volume in the phase of our interest and in a binary system $N_A + N_B = 1$.

From **Eqs. 6 and 12**, the intrinsic fluxes can be expressed as

$$J_A = -\frac{1}{2tV_m}\left[ N_A^+ \int_{x^{-\infty}}^{x_K} Y_A dx - N_A^- \int_{x_K}^{x^{+\infty}} (1-Y_A) dx \right] \qquad (12c)$$

$$J_B = \frac{1}{2tV_m}\left[ N_B^+ \int_{x^{-\infty}}^{x_K} Y_A dx - N_B^- \int_{x_K}^{x^{+\infty}} (1-Y_A) dx \right] \qquad (12d)$$

Since A diffuses right to left, $J_A = -D_A \frac{dC_A}{dx}$ and B diffuses left to right, $J_B = D_B \frac{dC_B}{dx}$. Intrinsic diffusion coefficients calculated at the Kirkendall marker plane as $D_A^K = 0.72 \times 10^{-14}$ and $D_B^K = 6.27 \times 10^{-14}$ $m^2/s$. Correctness of the calculations can be checked from the relation between the interdiffusion and intrinsic diffusion coefficients [**19**]

$$\tilde{D}^K = N_A^K D_B^K + N_B^K D_A^K \tag{13}$$

## 3.2 Pseudobinary approach in a ternary system

Let us now consider a ternary system of A-B-C, as shown in **Fig. 2.** For the sake of explanation, let us first consider that the homogeneity range of the γ phase does not change because of the addition of C. It further indicates that the element C replaces the element B. Now we need to select the end member compositions, such that the composition range of the elements A and B are the same with a fixed composition of element C in both. It can be any composition range; however, to compare with the results already calculated in the binary case, we consider the same composition difference of 0.2 for A and B between the end member alloys. We choose a fixed composition C, for example, 0.1. So the composition of the end members will be $A_{0.15}B_{0.75}C_{0.10}$ and $A_{0.35}B_{0.55}C_{0.10}$, since C replaces B. In most of the cases, the product phases also will grow with the same fixed percentage of C. For example, in (Si,Ge)/Mg diffusion couple, only the product phase $Mg_2$(Si,Ge) grows with the same ratio of (Ge,Si) alloy, since the product phase grows from this end member [**37**]. In many cases, when compositions of one or two elements are the same in both the end members, these do not vary in the interdiffusion zone also [**29, 38, 39**] (of course, if there is no uphill diffusion of any of the fixed elements, which will be discussed later). Then, it can be seen as an interdiffusion of A and B in the presence of other elements.

The diffusion rate of the elements, however, can greatly be affected because of the presence of other elements since defect concentrations (both thermal vacancies and structural antisites) [**40, 41**] and the thermodynamic driving forces in the γ product phase might change. Because of the change in diffusion rates, layer thickness of the product phase also will change. To explain and validate our calculations, we consider two cases. In one case, we assume that there is no change in layer thickness and the position of the Kirkendall marker plane is the same. This is considered first to validate the calculations. Then we consider the change in both in layer thickness and the position of the marker plane. These are shown in **Figs. 3a and 4a**.

We maintained the composition range of the elements of interest the same as binary since we wanted to validate our results. If no difference is seen in the layer thickness in the ternary system, the diffusion rates also should be the same. Since mainly two elements diffuse and the third element does not have the concentration gradient, it can be considered as a pseudobinary system. Moreover, it should be the same when we consider element A or B for our calculations. It can be seen immediately from $N_A$ vs. $x$ or $Y_A$ vs. $x$ plots in **Figs. 3a and c**. At the location of 40 μm, the interdiffusion coefficient calculated is the same as that calculated in the binary system. However, note that the $N_B$ is 0.66 (since $N_A = 0.24$ and $N_C = 0.10$) in the ternary system instead of 0.76 in the binary system. Since in the ternary system $N_A+N_B+N_C = 1$ and C replaces B, we rather measure the interdiffusion coefficients at $N_{B+C}$ (=$N_B+N_C$) = 0.76. For the interdiffusion coefficient calculations we could consider $N_B$ or $N_{B+C}$, since the composition is normalized and the composition range of the diffusion couple is the same. However, for the calculation of intrinsic diffusion coefficients, we need to consider the total composition of B and C, that is $N_{B+C}$. If we neglect C, then the total mol fraction (or the atomic fraction) will not be equal to one. So the calculation of intrinsic diffusion coefficients will not be proper, since we need to use the composition of the end members also (Eq.12). On the other hand, the composition profile for the element A is the same. So if we are interested in calculating the intrinsic diffusion coefficients also, we should calculate the diffusion parameter with respect to $N_A$ vs. $x$ or $N_{B+C}$ vs. $x$, as shown in **Figs. 3a and b**. Hence, we need to consider $Y_A$ vs. $x$ or $Y_{B+C}$ vs. $x$. This argument could be understood from the Eqs. 8, where the interdiffusion and intrinsic flux of one element is zero. We show the calculation of intrinsic diffusion coefficients with respect to $Y_A$ vs. $x$ plot, as shown in **Fig. 3c** following

$$D_A = \frac{1}{2t}\left(\frac{dx}{dN_A}\right)_K \left[ N_A^+ \int_{x^{-\infty}}^{x_K} Y_A dx - N_A^- \int_{x_K}^{x^{+\infty}} (1-Y_A)dx \right] \quad (12a)$$

$$D_B = -\frac{1}{2t}\left(\frac{dx}{dN_A}\right)_K \left[ N_{B+C}^+ \int_{x^{-\infty}}^{x_K} Y_A dx - N_{B+C}^- \int_{x_K}^{x^{+\infty}} (1-Y_A)dx \right] \quad (12b)$$

We considered exactly the same location of the marker plane and we get the same values of $D_A$ and $D_B$ as calculated in the binary system.

To validate the results, we considered the same layer thickness. However, mostly the layer thickness and even the location of the marker plane will actually change because of addition of another element, since defects concentrations and thermodynamic driving force might change. It could increase or decrease the growth rate depending on the system. Let us consider that the growth rate in the presence of C is lower and the layer thickness after annealing for 25 h is 90 µm. There is also change in the location of the marker plane and is found at 0.245 of $N_A$. So the location of the marker plane is at 40.5 µm from the alloy 3/γ phase interface, as shown in **Fig. 4a**. $Y_A$ vs. $x$ plot is shown in **Fig. 4c**. The interdiffusion and intrinsic diffusion coefficients calculated at the marker plane as $\tilde{D} = 1.68 \times 10^{-14}$, $D_A = 1.13 \times 10^{-14}$ and $D_B = 3.38 \times 10^{-14}$ m²/s.

So now we have seen that if experiments are conducted in a particular way, the ternary system can be treated as pseudobinary system and the same relations, developed for the binary system, can be used. We have shown the calculations at one particular composition only, however, like in a binary system, we can calculate the variation of the interdiffusion coefficients with compositions.

There could be a source of error because of the deviation/wrong estimation of the average composition in the ternary alloys. Few elements, such as Ti, Al evaporate during the melting and producing the alloys with very close to the average composition could be tiresome. This is also not a problem, if we can measure the composition correctly after melting. It is possible, if the alloy has only one phase and properly homogeneized. However, in a two phase alloy, it is not an easy task since error will be introduced during estimation of volume fractions from the area mapping for phase fractions to calculate the average compositions. Often, it is rather reliable to weigh the total amount before and after melting and calculation of the average composition can be done after deducting the weight loss from an element which evaporates during annealing. So it is important to examine the error one may introduce because of a wrong estimation of the end member compositions. For the sake of systematic calculation again, we consider that the composition of element C is more or less the same in both the end members. Further, we take into account the error only in one alloy (left hand side of the couple). We consider two examples, where the errors are 0.01 and 0.02 atomic fractions, respectively. So if the composition A is 0.16 and 0.17, as shown in **Figs 5a** and **5c,** the composition of (B+C) is 0.84 and 0.83. We calculate the interdiffusion coefficients at 40.5 mm from the alloy 5 and 6/$\gamma$ phase, after converting them to $Y_A$ vs. $x$ plots as shown in **Fig. 5b and d** as $1.55 \times 10^{-14}$ and $1.42 \times 10^{-14}$ m$^2$/s, compared to the value of $1.68 \times 10^{-14}$ when there was no error. So it indicates that the error should be kept within few percentages, since otherwise error in calculation could be high. Previously, we mentioned that one can calculate the interdiffusion coefficients even by considering just the composition profile of B instead of C. A source of error in such case could be if the composition of A is not different from the average but there is deviation in the composition of B and C. One will find two different values of interdiffusion coefficients if calculated with respect to composition profiles of A and B. Calculation from the profile of A will not give any error but the calculation from the profile of B will introduce error. On the other hand if the composition profile of B+C is considered then this error could be suppressed. So it is always safe to calculate with respect to element A or the element which does not have deviation from the average value. Depending on the condition, one should consider the composition profile for the calculation.

One can even calculate the tracer diffusion coefficients using the thermodynamic factor following a similar approach as it is explained for a binary system [**19**]. Since one element does not diffuse, the intrinsic diffusion coefficients can be related to the tracer diffusion coefficients ($D_i^*$) as

$$D_A = D_A^* \frac{d \ln a_A}{d \ln N_A}(1+W_A)$$

$$D_B = D_B^* \frac{d \ln a_B}{d \ln N_B}(1-W_B),$$

where $\frac{d \ln a_A}{d \ln N_A}$ and $\frac{d \ln a_B}{d \ln N_B}$ are the thermodynamic factors. $W_A = \frac{2N_A(D_A^* - D_B^*)}{M_o(N_A D_A^* + N_{B+C} D_B^*)}$ and

$W_B = \frac{2N_{B+C}(D_A^* - D_B^*)}{M_o(N_A D_A^* + N_{B+C} D_B^*)}$ are the vacancy wind effects and Mo is the structure factor. Note here that, we need to consider the total composition of (B+C) to estimate the vacancy wind effect of B.

### 3.3 Pseudobinary approach in a multicomponent system

As explained, we have chosen the alloy compositions such that the composition of one element is the same and add a third element at the cost of the second element. Similarly we can add many elements, which replace element B and then the calculation procedure will be the same. The interdiffusion coefficients can be calculated with respect to the composition profile of A or B. However, for the calculation of the intrinsic diffusion coefficients, we should consider the composition profile of A or the total of other elements (B+X). It should be noted here that we followed this procedure strictly to validate our calculations by comparing the results in multicomponent and the binary systems. Further, this systematic approach is useful to study diffusion of intermetallic compounds, where in general, one alloying element replaces one particular element. Otherwise, in order to study multicomponent diffusion following pseudobinary approach, the conditions should be followed such that the composition range of two elements of interest are the same and composition of all other elements should be the same in both the alloys to be bonded. For example, if we are interested to study interdiffusion in random solid solutions and calculate the interdiffusion at the equiatomic compositions (let say a 5 elements system), then only two elements, let say A and B could vary, for example, in the range of 0.15-0.25. 0.2 atomic fractions of all other elements should be added in both the alloys to be bonded, as shown in **Fig. 6**. Moreover, interdiffusion coefficients can be calculated at 0.2 atomic fractions from any of the composition profiles (A or B). However, the main difficulty in this system is that we cannot determine the intrinsic diffusion coefficients, since we cannot decide how to add the composition of other elements with A and B. Similar problem will be faced if one particular element replaces both the elements. In such case also only the interdiffusion coefficient can be calculated.

**3.4 Calculation of diffusion parameters in line compounds following the pseudobinary approach**

In Section 3.2 and 3.3, we explained the calculation of interdiffusion and intrinsic diffusion coefficients, which has wide homogeneity range, and that we can determine the composition gradient. However, many compounds grow with very narrow homogeneity range and as proposed by Wagner [**42**], the integrated diffusion coefficients, $\tilde{D}_{int}$ should be calculated. This is basically the interdiffusion coefficient integrated over the unknown small composition range of the phase let say, β. It can be expressed as

$$\tilde{D}_{int}^{\beta} = \int_{N_A^{'}}^{N_A^{''}} \tilde{D} dN_A \tag{13}$$

where $\Delta N_A^{\beta} = N_A^{''} - N_A^{'}$ is the composition range of the β phase. From **Eq. 13** and the Fick's first law expressed in **Eq. 3** we can write

$$\tilde{D}_{int}^{\beta} = \int_{x_I}^{x_{II}} \tilde{J} V_m^{\beta} dx = \tilde{J} V_m^{\beta} \Delta x^{\beta}, \tag{14}$$

since flux is more or less constant at any composition (or the position) inside the product phase and the molar volume of the phase considered is almost the same. $\Delta x^{\beta}$ is the thickness of the phase layer, where $x^I$ and $x^{II}$ are phase boundary positions in the interdiffusion zone, as shown in **Fig. 7**. Therefore, from **Eq. 11** and **14**, we can write

$$\tilde{D}_{int}^{\beta} = \frac{\Delta N_A \Delta x^{\beta}}{2t} \left[ (1-Y_A) \int_{x^{-\infty}}^{x^{*}} Y_A dx + Y_A \int_{x^{*}}^{x^{+\infty}} (1-Y_A) dx \right] \tag{15}$$

Note here that $\Delta N_A$ is the composition difference of the end members. It can be measured at any location inside the phase layer and the value will be the same. Similarly, as explained before for the phase with wide homogeneity range, the integrated diffusion coefficients can be calculated if the phase has very narrow homogeneity range. There is an added complication to determine the intrinsic diffusion coefficients, since we cannot determine the composition gradient. However, we can always calculate the ratio of diffusivities (for example in a ternary system) by taking the ratio of the **Eqs. 12a and 12b**

$$\frac{D_A}{D_B} = \frac{\left[ N_A^+ \int_{x^{-\infty}}^{x_K} Y_A dx - N_A^- \int_{x_K}^{x^{+\infty}} (1-Y_A) dx \right]}{\left[ -N_{B+C}^+ \int_{x^{-\infty}}^{x_K} Y_A dx + N_{B+C}^- \int_{x_K}^{x^{+\infty}} (1-Y_A) dx \right]} \tag{16}$$

Let us consider a diffusion couple between alloy 7: $A_{0.15}B_{0.75}C_{0.1}$ and alloy 8: $A_{0.35}B_{0.55}C_{0.1}$, where one line compound grows with composition $A_{0.25}B_{0.65}C_{0.1}$, as shown in **Fig. 7.** Suppose the marker plane is located at the distance of 40.5 μm from the alloy 5/β phase interface. The values are calculated as $\tilde{D}_{int}$ = 2.25 x $10^{-14}$ m²/s and $\frac{D_B}{D_A} = 2.33$.

## 4. Concluding remarks

Most of the diffusion studies available to date are on binary systems since it is relatively easy to calculate the diffusion parameters for binary systems. From the composition profile of one diffusion couple, we can calculate the interdiffusion coefficients for the compositions of all elements in a system. The intrinsic diffusion coefficients can be calculated for the composition of the marker plane. Limited studies are available for ternary systems because of the difficulties in setting up and conducting experiments on ternary systems. The interdiffusion coefficients for ternary systems can only be calculated for a single composition, where the profiles for two different diffusion couples intersect. Intrinsic diffusion coefficients for ternary systems are extremely difficult to calculate since Kirkendall markers should be present at that composition. It is almost impossible to predict the compositions of the end members of the diffusion couples such that Kirkendall markers are present there. Dayananda *et al*.

studied interdiffusion in the Ni-Cr-Co-Mo [28] and Cu-Ni-Zn-Mn [29] quaternary systems. They calculated the main interdiffusion coefficients in the Ni-Cr-Co-Mo system when two elements did not develop any diffusion profile because the compositions of these elements in both end members of the diffusion couples were more or less the same. Similarly, interdiffusion in the Cu-Ni-Zn-Mn system was explained only based on the calculation of flux. We have developed a method of calculating the variation in interdiffusion coefficients with varying composition in the interdiffusion zone. Our method can even be used to calculate the intrinsic diffusion coefficients from only one composition profile for multicomponent pseudobinary systems. It is necessary to select the composition profiles such that only two elements diffuse into the interdiffusion zone and that the compositions of all other elements are constant throughout the diffusion couple. The added advantage of this method can readily be understood. In previously available methods, it was necessary to select end member compositions such that the composition profiles intersected. Since the points of intersection cannot be predicted, it is difficult to systematize the study in order to determine the effect of the composition on diffusion-controlled growth of phase layers or on diffusion parameters. In our method, on the other hand, one can easily examine the effect of the change in the composition of element C on diffusion-controlled growth of phase layers or on diffusion parameters by measuring the thickness of either the phase layer or the interdiffusion zone. The diffusion coefficients can subsequently be calculated based on the variation in the composition of element A or B. From the few diffusion-couple experiments in which the compositions of C were fixed, one can systematically study the effect of the change in the composition of C on diffusion-controlled growth of phase layers or on diffusion parameters. Further, another element, D, can also be added with C to study the diffusion of A and B in the presence of C and D. Requirements for further improvements can be explained based on the composition profiles for the Ni-Pt-Al system, as shown in **Fig. 8** [39]. In this study, diffusion couples were prepared with a fixed Pt concentration to determine the role of Pt on the growth and diffusion in the $\gamma'$-(NiPt)$_3$Al intermetallic compound and the $\gamma$-Ni(PtAl) solid solution phase. The thickness of both phase layers increased with increasing Pt concentration. The average interdiffusion diffusion coefficients for the elements were subsequently calculated. The integrated diffusion coefficients should first be calculated based on **Eq. 14** and then should be divided by the homogeneity range of the phase to calculate the average interdiffusion coefficients. Two elements of interest can have different composition ranges (because the alloys were not prepared with the exact compositions desired) in some diffusion couples, and two different average interdiffusion coefficients can be calculated for the respective elements. The diffusion rates of two elements could even be affected differently. The composition ranges of the phases vary with increasing Pt concentration, and various composition ranges used to calculate the average interdiffusion coefficients for various Pt concentrations could be a source of error. We of course realize that instead of calculating the average interdiffusion coefficients, it is possible to improve the calculation if the method described in this manuscript is followed. The amount of error in the calculation would decrease since instead of calculating the average interdiffusion coefficients, we would calculate the interdiffusion coefficients for various compositions of the elements. Moreover, as was already explained, the interdiffusion coefficients should be calculated from the composition profiles for Al or (Ni + Pt) to minimize error. Further improvements in the experiments are required in order to apply our method to calculating the interdiffusion coefficients for this pseudobinary system. The compositions of the end members of the diffusion couples should be selected such that only one phase grows into the interdiffusion zone. In this way, deviation of the composition profile from the straight line connecting the compositions of the end members can be prevented since the composition range will be relatively small. Experiments are currently being conducted in our laboratory to satisfy these conditions. It should be noted that the proposed model will fail if any component whose composition is fixed in both end members diffuses uphill in a multicomponent system. Diffusion will occur under this condition, and the composition of a particular element will deviate from the average fixed composition of the element even if its composition is the same in both end members of the diffusion couple. If the product phase is a line compound, however, this method will be rather straightforward to use, and uphill diffusion will rarely occur. Fukaya et al. [**43**] studied interdiffusion in the $Ni_3Al[X = V, Ti, Nb]$ system, where Ni23Al/Ni23Al2X alloys were coupled. They mentioned it as a pseudobinary system. However, Al+X was not kept constant in both the couples. Even in some cases, Al uphill diffusion was evident from the composition profile. Although only 2 at.% of X was used, ideally the average composition should be Ni21Al2X instead of Ni23Al2X since X replaces Al in this alloy.

**Acknowledgement:** The authors acknowledge the financial support from the Department of Science and Technology, India, which enabled them to carry out this research.

**Reference**

[1] A. Paul, A.A. Kodentsov and F.J.J. van Loo, Journal of Alloys and Compounds 403 (2005) p. 147-153.

[2] V.D. Divya, U. Ramamurty and A. Paul, Journal of Materials Research 26 (2011) p. 384-2393.

[3] S. Santra and A. Paul, Philosophical Magazine Letters 92 (2012) p. 373-383.

[4] S.S.K. Balam, H.Q. Dong, T. Laurila, V.Vuorinen, A. Paul, Metallurgical Materials Transactions A 42 (2011) p. 1727-1731.

[5] A.K. Kumar, T. Laurila, V. Vuorinen and A. Paul, Scripta Materialia 60 (2009) p. 377-380.

[6] A. Paul, A.A. Kodentsov and F.J.J. van Loo, Journal of Alloys and Compounds 403 (2005) p. 147-153.

[7] C. Cserháti, A. Paul, A.A. Kodentsov, M.J.H. van Dal, F.J.J. van Loo, Intermetallics 11 (2003) p. 291-297.

[8] S. Prasad and A. Paul, Intermetallics 19 (2011) p. 1191-1200.

[9] S. Roy and A. Paul, Philosophical Magazine 92 (2012) p. 4215-4229.

[10] S. Prasad and A. Paul, Defects Diffusion Forum 323-325 (2012) p. 459-464.

[11] S. Prasad and A. Paul, Journal of Phase Equilibria and Diffusion 32 (2011) p. 212-218

[12] S. Prasad and A. Paul, Defects and Diffusion Forum 312-315 (2011) p. 731-736

[13] S. Prasad and A. Paul, Acta Materialia 59 (2011) p. 1577-1585.

[14] A. Paul , C. Ghosh and W.J. Boettinger, Metallurgical and Materials Transactions A 42A (2011) p. 952-963.

[15] S.S.K. Balam, and A. Paul, Journal of Materials Science 46 (2011) p. 889-895.

[16] S.S.K. Balam, and A. Paul, Metallurgical and Materials Transactions A, 41A (2010) p. 2175-2179.

[17] A. Paul, A.A. Kodentsov, G. de With, F.J.J. van Loo, Intermetallics 11 (2003) p. 1195-1203.

[18] R. Bouchet and R. Mevrel, Acta Materialia 50 (2002) p. 4887-4900

[19] F.J.J. van Loo, Progress in Solid State Chemistry 20 (1990) p. 47-99.

[20] A. Paul, Journal of Materials Science: Materials in Electronics 22 (2011) p. 833-837

[21] A. Paul, The Kirkendall effect in solid state diffusion, PhD Thesis, 2004, Eindhoven University of Technology, Eindhoven, The Netherlands.

[22] C. Ghosh and A. Paul, Acta Materialia 57 (2009) p. 493-502.

[23] C. Ghosh and A. Paul, Intermetallics 16 (2008) p. 955-961.

[24] C. Ghosh and A. Paul, Acta Materialia 55 (2007) p. 1927-1939.

[25] A. Paul, A.A. Kodentsov and F.J.J. van Loo, Intermetallics 14 (2006) p. 1428-1432.

[26] A. Paul, M.J.H. van Dal, A.A. Kodentsov and F.J.J. van Loo, Acta Materialia 52 (2004) p. 623-630.

[27] Y.H. Sohn and M.A. Dayananda, Metallurgical and Materials Transactions 33A (2002) p. 3375-3392.

[28] K.E. Kansky and M.A. Dayananda, Metallurgical and Materials Transactions 16A (1985) p. 1123-1132.

[29] Y. Heaney and M.A. Dayananda, Metallurgical and Materials Transactions 17A (1986) p. 983-990.

[30] M.A. Dayananda, Editors: M.A. Dayananda and G.E. Murch, Diffusion in Solids: recent developments, A publication of the Metallurgical Society, Inc., Pennsylvania, USA, 1985.

[31] S. Krishtal, A.P. Mokrov, A.V. Akimov, and P.N. Zakharov: Frz. Metal. Metalloved. 35 (1973) p. 1234-40.

[32] M.S. Thompson and J. E. Morral, Acta Metallurgica 34 (1986) p. 2201-03.

[33] M.K. Stalker, J.E. Morral and A.D. Romig, Metallurgical Transactions A, 23 (1992) p.3245.

[34] Y. Minamino, Y. Koizumi, N. Tsuji, T. Yamada and T. Takahashi, Materials Transactions 44 (2003) p. 63-71.

[35] T. Takahashi, K. Hisayuki, T. Yamane, Y. Minamino and T. Hino, Materials Transactions 44 (2003) p. 2252 – 2257.

[36] M.K. Stalker, J.E. Morral and A.D. Romig, Metallurgical Transactions A 23A (1992) 3245-49.

[37] Y. Mizuyoshi, R. Yamada, T. Ohishi, Y. Saito, T. Koyama, Y. Hayakawa, T. Matsuyama, H. Tatsuoka, Thin solid Films 508 (2006) p. 70-73.

[38] K.N. Kulkarni, B. Tryon, T.M. Pollock and M.A. Dayananda, Journal of Phase Equilibria Diffusion 28 (2007) p. 503-509.

[39] V.D. Divya, U. Ramamurty and A. Paul, Philosophical Magazine 92 (2012) p. 2187-2214.

[40] W. Li, H. Yang, A. Shan, L. Zhang, J. Wu. Intermetallics 14 (2006) p. 392-395.

[41] S. Prasad and A. Paul, Intermetallics 22 (2012) p. 210-217.

[42] C. Wagner, Acta Metallurgica 17 (1969) p. 99-107.

[43] H. Fukaya, Md. Moniruzzaman, Y. Murata, M. Morinaga, T. Koyama, W. Hashimot, K. Tanaka and H. Inui, Defects and Diffusion Forum 297-301 (2101) 384-389

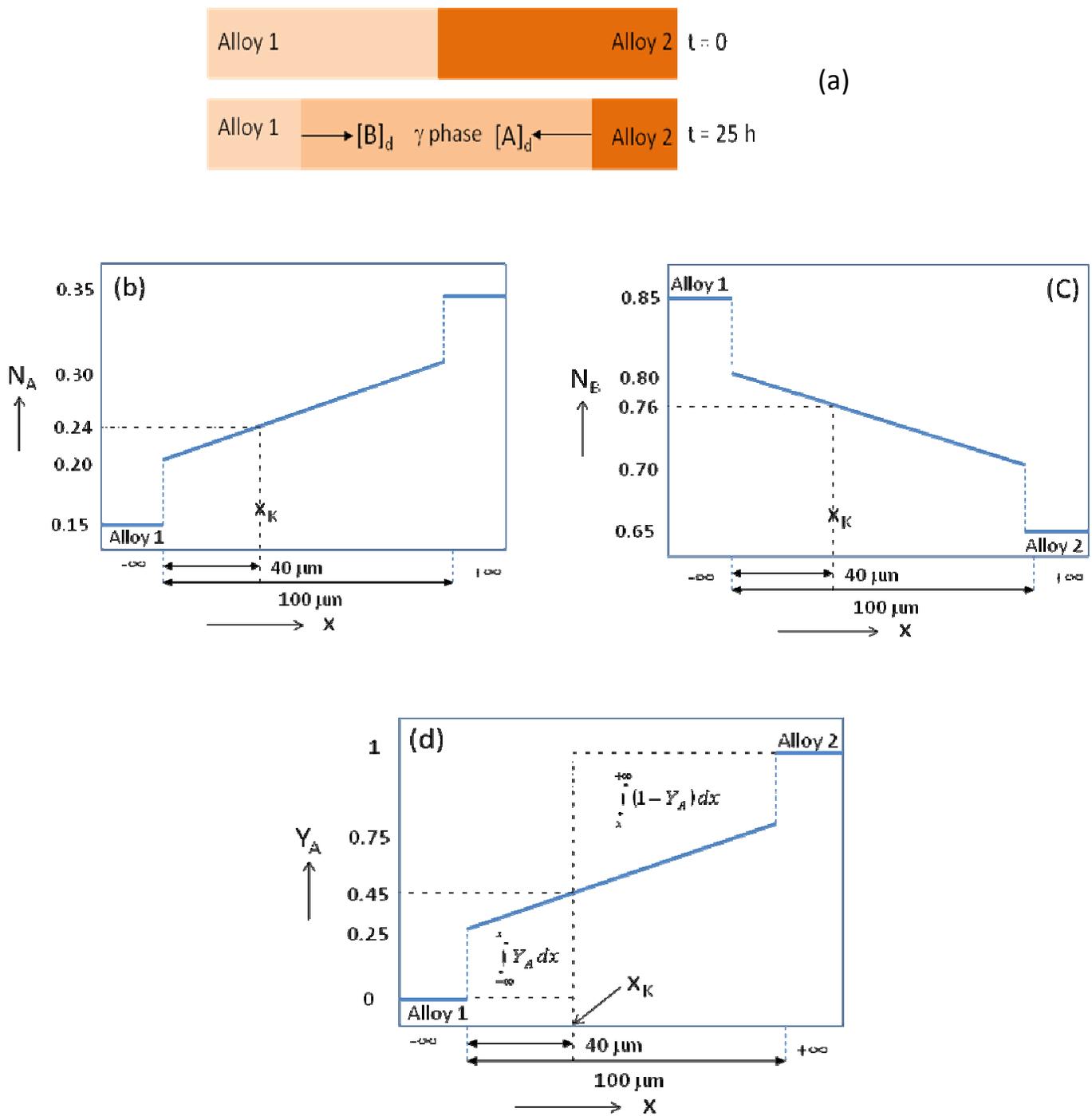

Fig. 1 (a) Diffusion couple of alloy 1 and 2, (b) composition profile with respect to element A (c) composition profile with respect to element B and (d) composition normalized variable $Y_A$ vs. x plot.

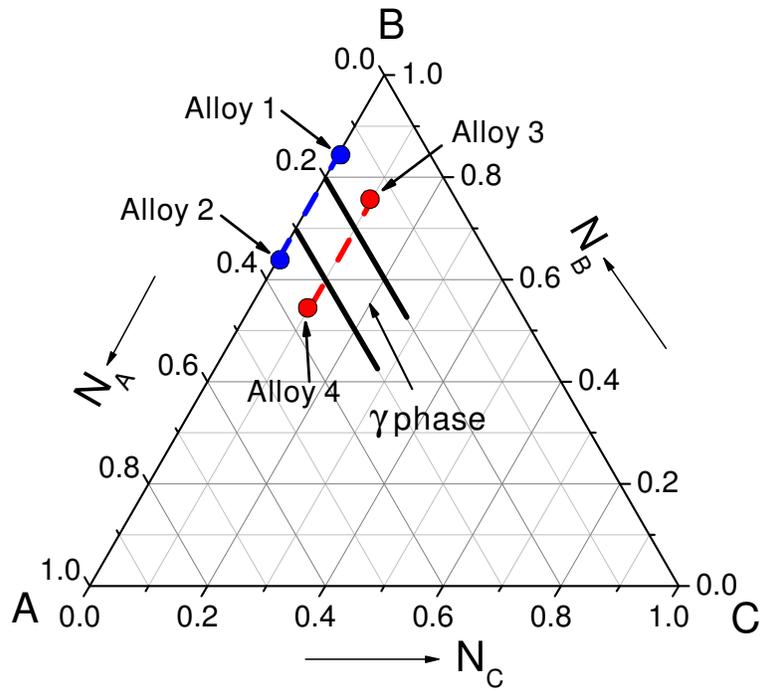

Fig. 2 Compositions of the alloys chosen for diffusion couples.

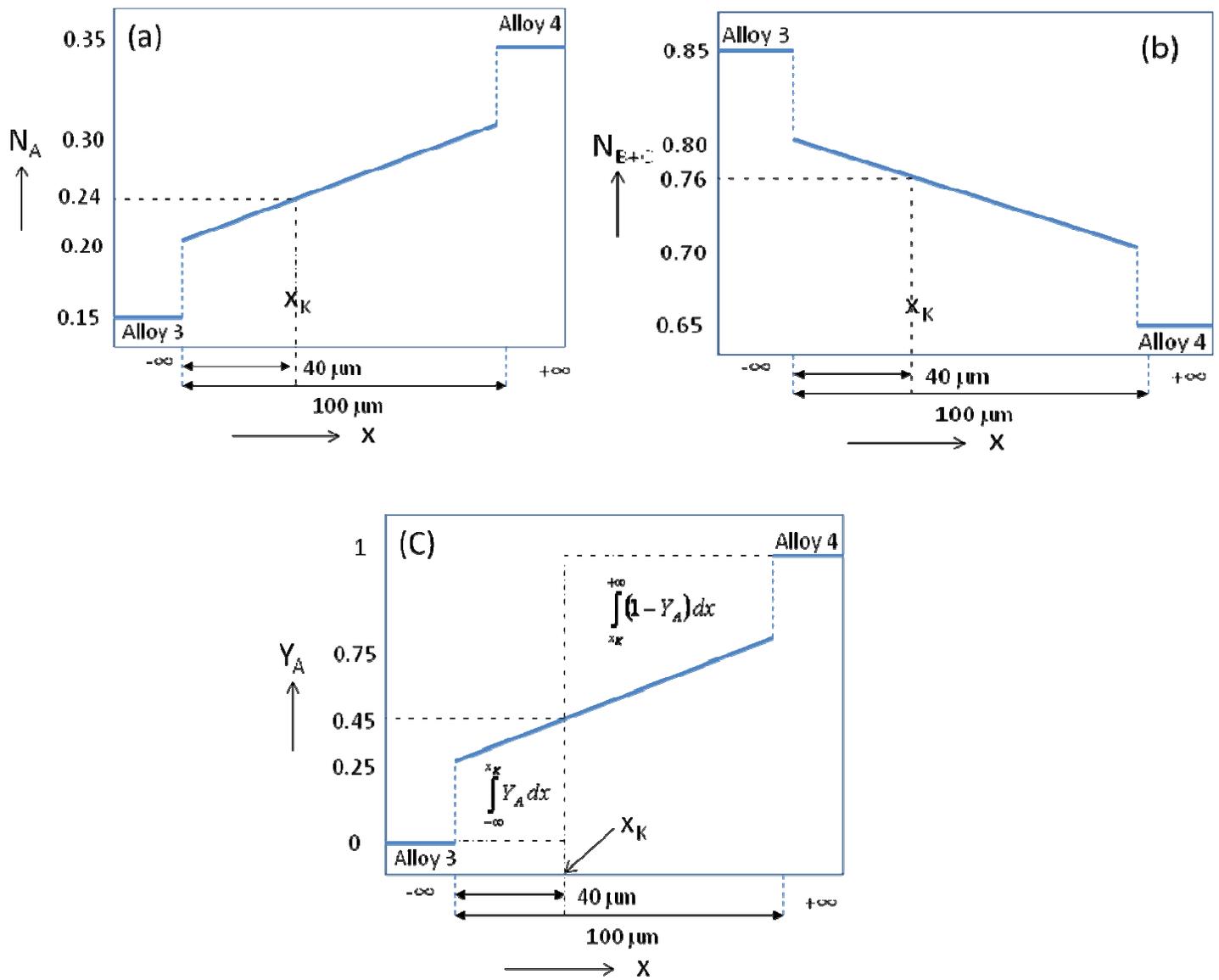

Fig. 3 Composition profiles of a diffusion couple in a ternary A-B-C system (a) composition profile of element A (b) composition profile of elements (B+C), (c) composition normalized variable $Y_A$ vs. x plot.

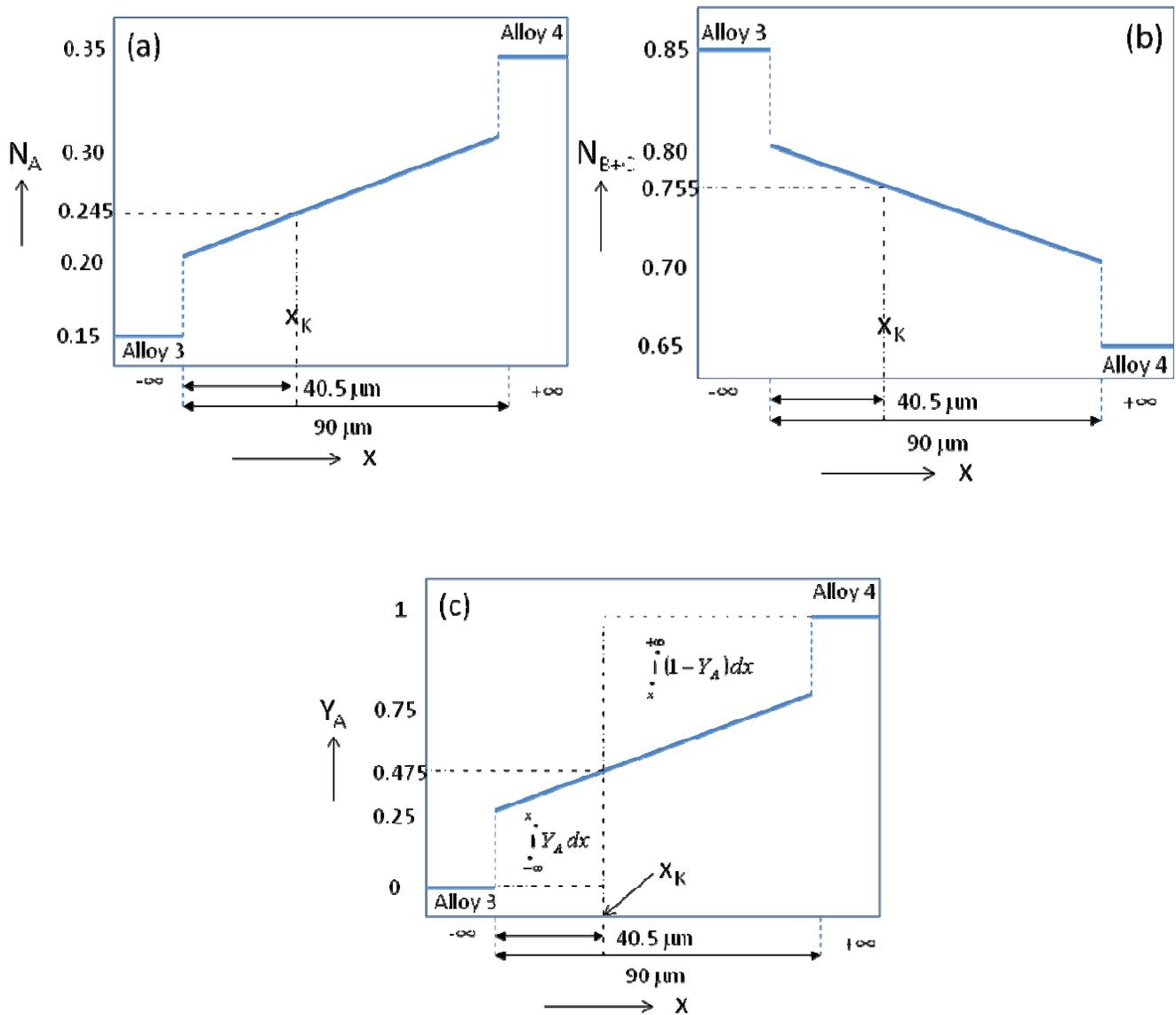

Fig. 4 Composition profiles of a diffusion couple in a ternary A-B-C system, where layer thickness is affected because of addition of C with A and B (a) composition profile of element A (b) composition profile of elements (B+C), (c) composition normalized variable $Y_A$ vs. x plot.

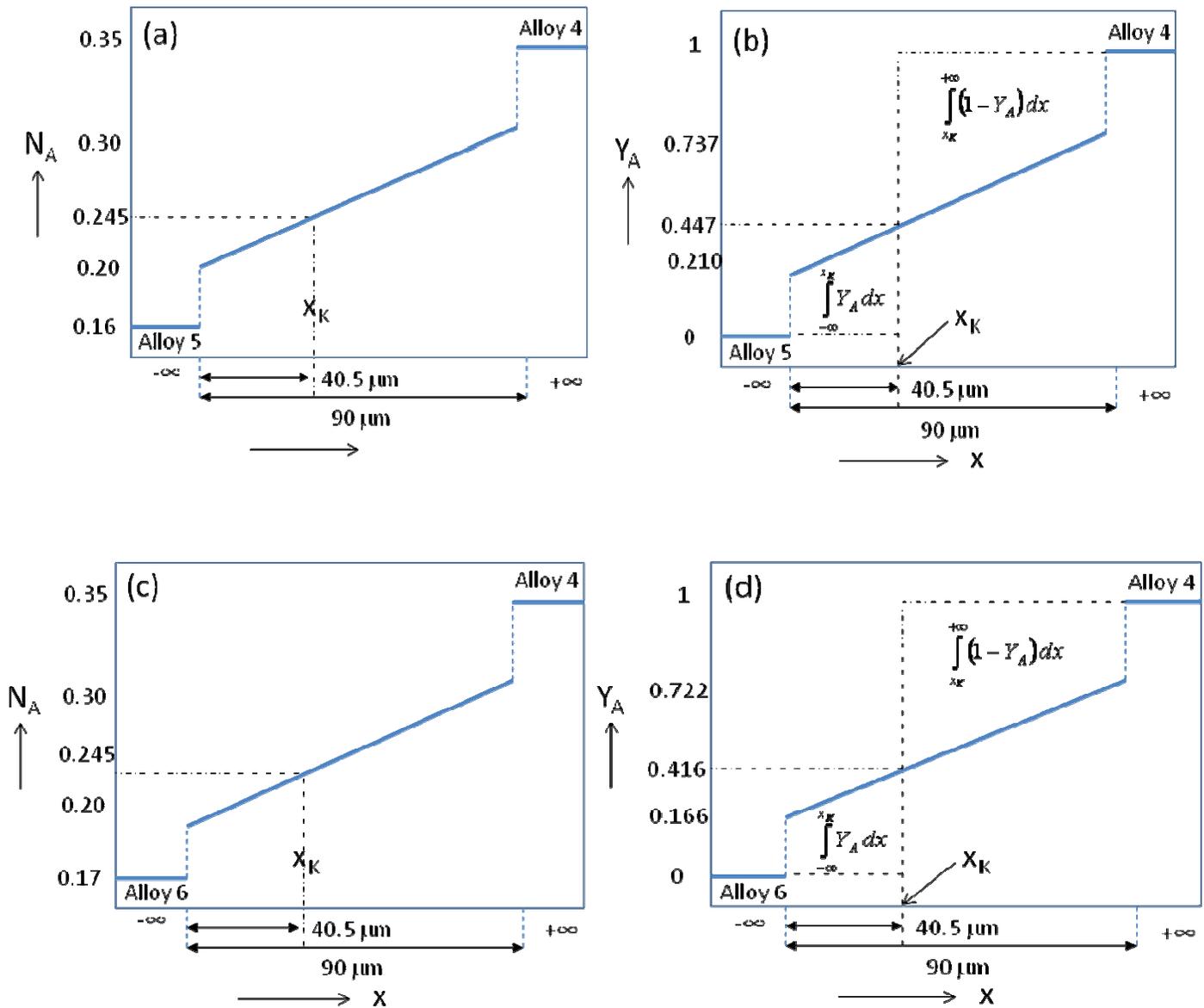

Fig. 5 (a) composition profile of element A in $A_{0.16}B_{0.74}C_{0.1}/A_{0.35}B_{0.55}C_{0.1}$ couple and (b) $Y_A$ vs. x plot of the same (c) composition profile of element A in $A_{0.17}B_{0.73}C_{0.1}/A_{0.35}B_{0.55}C_{0.1}$ couple and (d) $Y_A$ vs. x plot of the same.

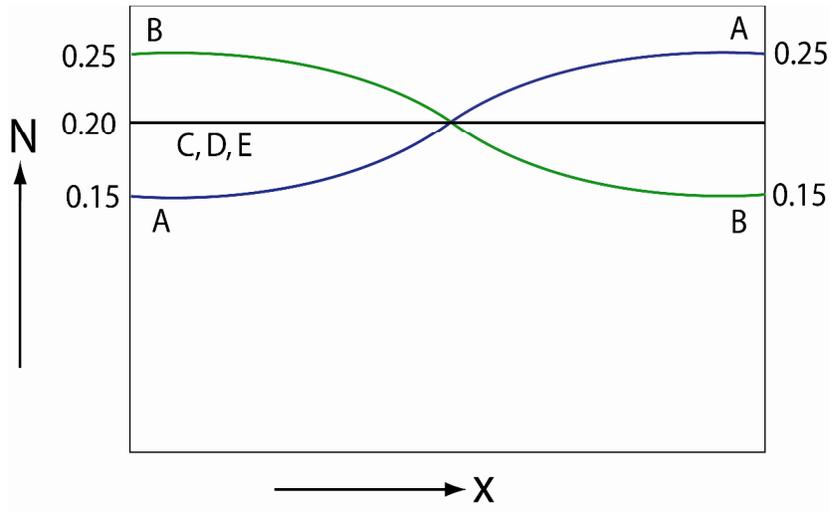

Fig. 6 Composition profile of multicomponent diffusion couple in random solid solution.

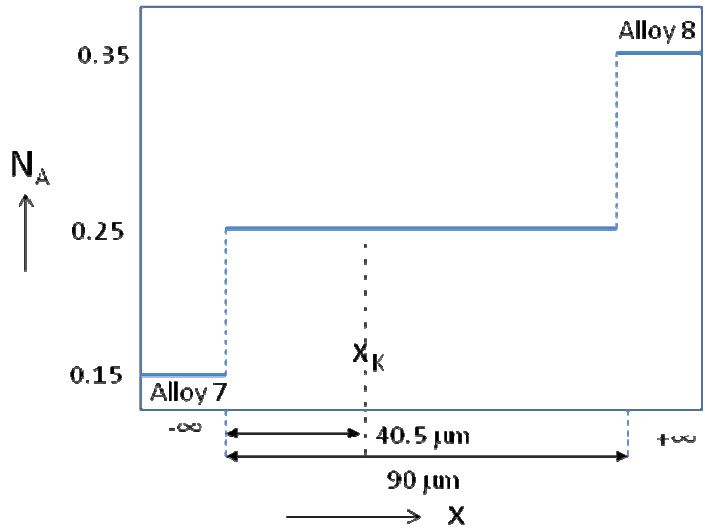

Fig. 7 Composition profile of element A in a multicomponent system, where a line compound grows in the interdiffusion zone.

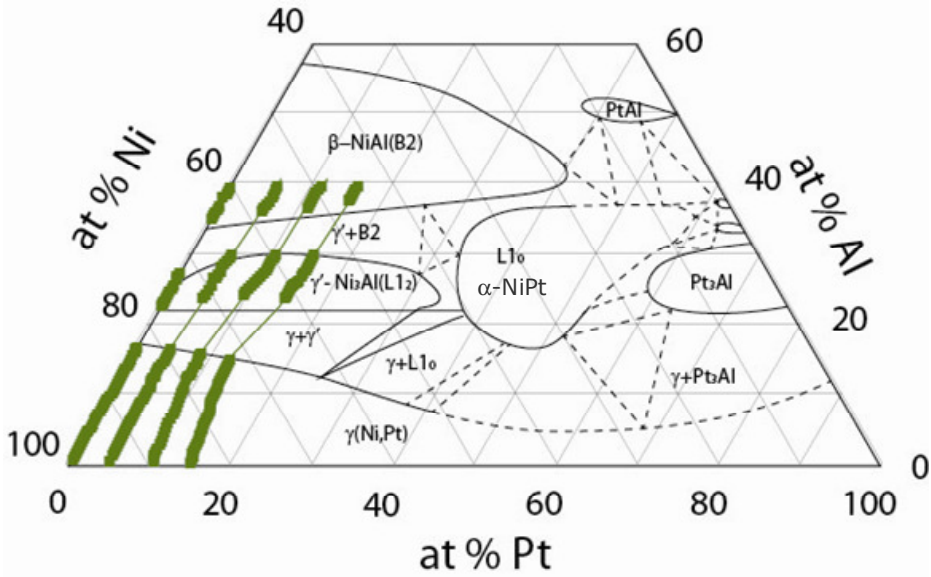

Fig. 8 Concentration profiles developed in the Ni/Ni40Al binary couple and in the Ni(XPt)/(Ni,XPt)40Al ternary couples (where X=5,10,15) are shown on the Ni-Al-Pt phase diagram [39].